# Confined Acoustic Phonon in CdS$_{1-x}$Se$_x$ Nanoparticles in Borosilicate Glass

Sanjeev K. Gupta, Prafulla K. Jha[1], Satyaprakash Sahoo[*], A. K. Arora[*] and Y. M. Azhniuk[#]

Department of Physics, Bhavnagar University, Bhavnagar- 364 022, India.
[*]Material Science Division, Indira Gandhi Centre for Atomic Research, Kalpakkam, Tamilnadu, India.
[#]Institute of Electron Physics, Ukr. Nat. Acad Sci., Universytetska St. 21, Uzhhorod 88000, Ukraine.

## ABSTRACT

We calculate low-frequency Raman scattering from the confined acoustic phonon modes of CdS$_{1-x}$Se$_x$ nanoparticles embedded in borosilicate glass. The calculation of the Raman scattering by acoustic phonons in nanoparticles has been performed by using third-order perturbation theory. The deformation potential approximation is used to describe the electron-phonon interaction. The Raman-Brillouin electronic density and the electron-phonon interaction are found to increases with decreasing size of nanoparticle. A good agreement between the calculated and reported low-frequency Raman spectra is found.

# I. INTRODUCTION

Semiconductor nanocrystals of a size comparable to or smaller than the exciton Bohr radius in bulk material have attracted considerable attention from the fundamental physics viewpoint and from the interest in the applications in functional devices,[1-3] because of their novel optical properties arising from the discrete electronic energy levels.[4-6] To understand the optical properties of semiconductor nanocrystals, it is absolutely necessary to consider the zero dimensional confinement effects on the electronic states in a system. Raman spectroscopy has emerged as a powerful and non-destructive technique for estimating the composition of glass embedded nanocrystals[6-8] and their size.[8-10] The size-quantized acoustic phonon modes were observed in the low-frequency Raman scattering, which provides an understanding of electron-phonon interactions. The low-frequency phonon modes (confined acoustic phonon) of nanoparticles bear a unique signature of their structural and mechanical properties besides modifying the electronic structure. There exist several studies on low-frequency Raman scattering from elastic spherical nanoparticle vibrating with frequencies inversely proportional to their size. Thus the question of the significance of phonon confinement in a theoretical description of Raman scattering from nanocrystals still remains open. As far as the theoretical description of the quantized acoustic vibrations is concerned, they have been mostly done by using the well known Lamb's model.[11] This model yields the spheroidal and torsional vibrational modes when the Navier equation is solved under the proper boundary condition of the particle surface.[11-14] The quantized vibrational modes are labeled by the angular momentum quantum number $l$ and the harmonic/branch number $n$. The spheroidal (SPH) $l=0$, $n=0$ and $l=2$, $n=0$ modes are Raman active and are observable in Raman experiment.[12] To consider the effect of matrix on the nanoparticle vibrations the complex frequency model (CFM)[7,13-15] is used which results in the frequency shift and damping of modes. This model has the drawback that the displacement fields are not orthonormalized required for the calculation of electron-phonon interaction and Raman line shape. To overcome this Murray and Saviot[16] proposed a model known as core shell model (CSM) to obtain the orthonormalized displacement. In this model the particle is considered as core while the matrix as shell and the mean square displacement $\langle u^2 \rangle_p$ of the nanoparticle is obtained as a function of frequency of the nanoparticle vibration.

Recently we have investigated the low-frequency Raman scattering from $CdS_{1-x}Se_x$ nanocrystals embedded in borosilicate glass arising from confined acoustic phonon.[17] In addition to the breathing mode ($l=0$, $n=0$: SPH) and quadrupolar mode ($l=2$, $n=0$: SPH), two additional modes are found in the spectra. Based on the ratio of the frequencies of the new modes to those of the quadrupolar mode calculated using theoretical approaches, the new modes were assigned to first overtone of the quadrupolar mode ($l=2$, $n=1$) and $l=1$, $n=0$ torsional mode. The appearance of the forbidden torsional mode[12] is attributed to the near-spherical shape of the nanoparticle found from high-resolution TEM[18]. To further confirm the appearance of torsional mode a detail study of polarized Raman is in progress. Though we have been able to predict the Raman shift and calculate the line shape from the mean square displacement $\langle u^2 \rangle_p$ using core-shell model, the calculation ignored the electron-phonon interaction. The mean square displacement is nothing but the measure of the internal motion of the nanoparticle. The Raman spectrum is governed by both $\langle u^2 \rangle_p$ and electron-phonon interaction element. Therefore, it is necessary to have theoretical Raman spectra that are determined by quantum confinement of electronic states, electron-phonon interaction and localized acoustic modes for a meaningful comparison with experimental Raman scattering.

In the present paper, we report the results of the calculation of electron-acoustic phonon interaction and Raman-Brillouin electronic density following the approach developed by Gupalov and Merkulov[19] and Huntizinger et al[20] and compare them with the reported experimental low-frequency Raman spectra[17] for the $CdS_{1-x}Se_x$ nanocrystals. The present calculation is done for the $l=0$, $n=0$: SPH mode. Size-quantization of both electron and hole states is treated within the envelope wave function approximation.[19,20] Deformation potential interaction between the confined electronic states and the acoustic vibrations is considered.

## II. THEORETICAL CONSIDERATIONS

The Raman intensity, under the framework of theory of resonance Raman scattering is calculated by using the third-order perturbation theory and expressed as,[19]

$$I(\omega) \propto \sum_{\nu/\omega_\nu = |\omega|} \left| A_\nu^{e(h)} \right|^2, \qquad (1)$$

where $A_\nu$ is the scattering amplitude from state $|i\rangle$ with an incident photon $\vec{k}_i$ to state $|f\rangle$ with a scattered photon $\vec{k}_f$ through emission or absorption of a phonon mode $\nu$ with energy $\hbar\nu$. In the Raman scattering process, both electron and hole states contributing to the electron-phonon interaction is to be considered as the intermediate electronic excitation can be either free or bounded electron-hole pairs. Moreover, in the strong confinement limit, the size quantized electron and hole have identical wave functions, and coupling to polar phonons vanishes. Coupling to acoustic phonons occurs through deformation potential and piezoelectric interactions. However, we consider only the deformation potential as this is dominant in the small size region[20] and the electron-phonon Hamiltonian for the deformation potential is expressed as,

$$H_{e-\nu}^{e(h)} = D_{e(h)}\left(\overline{\nabla}.\overline{u}_q\right), \tag{2}$$

where $D_{e(h)}$ is the conduction (valance) hydrostatic deformation-potential energy and $\overline{u}_q = \sqrt{\hbar/2\omega_\nu}\, \overline{u}_q(\overline{r})$ with $\overline{u}_q(r)$ as displacement field. The scattering amplitude in Eq. (1) is due to both electron and hole contributions and can be expressed as,[19,20]

$$A_\nu^{e(h)} = R\int d^3 r \rho^{e(h)}(\overline{r})\left[H_{e-\nu}^{e(h)}(\nu)\right](\overline{r}), \tag{3}$$

where $R = \sum_\gamma R_\gamma^{em} R_\gamma^{abs}$ is a Resonance factor and $\rho^{e(h)}(\overline{r})$ the Raman-Brillouin electronic density (RBED) defined as,

$$\rho^{e(h)}(\overline{r}) = \frac{1}{R} \sum_{\{\beta,\alpha / \beta^{h(e)} = \alpha^{h(e)}\}} R_\beta^{em} R_\alpha^{abs} \psi_\beta^{e(h)}(\overline{r})^* \psi_\alpha^{e(h)}(\overline{r}), \tag{4}$$

where $\psi_\alpha^{e(h)}(\overline{r})$ are the envelope wave functions for the electron (hole) part of the state $|\alpha\rangle$, $\rho^{e(h)}(\overline{r})$ is simply a linear combination of joint local electronic densities $\psi_\beta^{e(h)}(\overline{r})^* \psi_\alpha^{e(h)}(\overline{r})$ weighted by the optical resonance factors associated with the photon emission and absorption can be written as, $R_\beta^{em} = \frac{\langle f|H_{e-r}|\beta\rangle}{(E_\beta + \hbar\omega - \varepsilon_i) - i\Gamma_\beta}$ and $R_\alpha^{abs} = \frac{\langle \alpha|H_{e-r}|i\rangle}{(E_\alpha - \varepsilon_i) - i\Gamma_\alpha}$ respectively. Where $E_{\alpha(\beta)}$ and $\Gamma_{\alpha(\beta)}$ are the energy and homogeneous broadening of the $\alpha$ and $\beta$ states, respectively. In the case of large conduction and valence band offsets between the nanoparticle and surrounding medium $\hbar\omega = \varepsilon_i - \varepsilon_f$ is the differences between incident and scattered photon energies. The RBED is a fundamental quantity for the

third order interaction process as all electronic states contributes to it. The RBED is the electronic density distribution that emits or absorbs a given vibrational mode through a Raman process. The electron and hole envelop wave function can be written as,[19]

$$\psi_{n,l,m}(\bar{r}) = \frac{1}{j_{l+1}(\eta_{n,l})}\sqrt{\frac{2}{a^3}} j_l\left(\eta_{n,l}\left(\frac{r}{a}\right)\right)\overline{Y}_l^m(\theta,\phi) \quad (5)$$

Here, $a$ is the nanoparticle radius, $\eta_{n,l}$ is the $n^{th}$ zero of the $j_l$ spherical Bessel function and this wave function would vanish at the boundary at $r=a$ and $\overline{Y}_l^m(\theta,\phi)$ is normalized spherical harmonics. Therefore by combining all terms the Raman intensity can be written as,[19]

$$I(\omega,a) \propto |R|^2 g(\omega)|n(-\omega)|\omega|M[q(\omega)a]|^2 \quad (6)$$

In Eq. (6), $g(\omega)$ is phonon density of states and $n(\omega) = \dfrac{1}{\exp\left(\hbar\omega/k_B T\right)-1}$, is the phonon occupation number with $T$ as temperature of phonon bath and $n(\omega)+1 = |n(-\omega)|$ due to overall energy conservation. The matrix element factor which determines the electron-phonon interaction for the isotropic displacement: $l=0$ in the calculation of Raman intensity is expressed as,

$$|M[q(\omega)a]| = \int dr\, r^2 \rho^{e(h)}(\bar{r}) j_0(qr) \quad (7)$$

### III. RESULTS AND DISCUSSION

Figure 1 presents the low-frequency experimental Raman spectra along with the calculated resonance Raman spectra of CdS$_x$Se$_{1-x}$ nanoparticles for two sizes. The Raman spectra for two sizes are displayed in two panels. The experimental spectra presented in the upper part of the figure show four peaks; two spheroidal modes, one torsional mode against the selection rule and one overtone of quadrupolar ($l$=2, $n$=0: SPH) mode. The lower part of the figure presents the calculated Raman spectra for $l$=0, $n$=0: SPH (breathing mode). The simulated spectra are reported only for the $l$=0 mode as for *s-s* electronic transition only the azimuthal quantum numbers are needed.[19] In the calculation of Raman spectra acoustic resonances are considered. It is seen from the Fig. 1 that there is a good agreement between the measured and simulated Raman spectra for $l$=0 mode for $CdS_xSe_{1-x}$ nanoparticle of both sizes. In the case of 4.6 nm $CdS_xSe_{1-x}$ nanoparticle the second peak at ~29 cm$^{-1}$ is the overtone of $l$=0, $n$=0 mode. The observed blue shift of the Raman peaks with decreasing nanoparticle size is due to both electronic and acoustic phonon confinement. The peak frequencies are almost inversely proportional to the particle diameter in agreement with the experimental spectra. The overtone mode is not plotted in the case of 2.2 nm size nanoparticle. The figure also reveals that the width of peak increases with decreasing nanoparticle sizes, which is due to the Raman-Brillouin electronic density (RBED) and electron-phonon interaction matrix element expressed in Eq. (4) and Eq. (7) respectively. However, in the present calculation, size distribution has not been considered which mainly contributes to the inhomogeneous broadening. The homogeneous broadening resulting mainly due to smaller size of the particle and its interaction with matrix has been well produced by the complex frequency approach. Further, the size distribution for the samples is not known.

Figure 2 presents the radial density distribution of the RBED inside a nanoparticle. The RBED is calculated for the homogeneous broadening $\Gamma$'=0.1. In this case the RBED, $\rho^{e(h)}(\bar{r})$ reduces to the electronic density associated with the first electron and hole states and the electronic density of states is discretazied.[19] The vibrational mode $l$=0 may interacts with the RBED and are responsible for the symmetric deformation potentials.[20] The Fig. 2 depicts that the Raman-Brillouin electronic density increases towards the centre of the particle and therefore is responsible for the increases of Raman linewidth observed in Fig. 1. The RBED inside a nanoparticle determines the electron-phonon coupling and therefore we have also calculated the electron-phonon matrix element. Figure 3 presents the electron-phonon interaction matrix element calculated for the $\Gamma'=0.1$. It is seen from the figure that the electron-phonon matrix element *i.e* the electron-phonon interaction increases with the reduction in size. The increase in the electron-phonon matrix element is more rapid below 8 nm. Consequently the overall scattering efficiency increases with decreasing $CdS_xSe_{1-x}$ nanoparticle size.

## IV. CONCLUSION

To summarise, low-frequency Raman spectra for *l*=0 confined acoustic phonon is calculated using third-order perturbation theory that takes into account the mean square amplitude, electron-phonon interaction and Raman-Brillouin electronic density. Both the electron-phonon interaction and the RBED are found to increase with decreasing particle size. A good agreement between simulated and experimental Raman spectra is found for pure radial vibrational mode i.e. *l*=0, *n*=0: spheroidal mode.

## ACKNOWLEDGEMENT

We thank DAE-BRNS, Mumbai for the financial assistance.

**Figures Caption**

1. Measured and simulated low-frequency Raman spectra of $CdS_xSe_{1-x}$ nanoparticle embedded in borosilicate glass (a) $CdS_xSe_{1-x}$ ($r$=2.2 nm), (b) $CdS_xSe_{1-x}$ ($r$=4.6 nm). Open circles represent the data and full curve is the fitted shape consisting of an exponential background and four Lorentzian components. The background and the four components are also shown in the figure. The lower panel of the figure presents the simulated Raman spectra for $l$=0, $n$=0 spheroidal mode.

2. Radial density distributions (RBED) in $CdS_xSe_{1-x}$ nanoparticles of sizes 2.2 and 4.6 nm.

3. Variation of electron-phonon matrix-element with size of the $CdS_xSe_{1-x}$ nanoparticle.

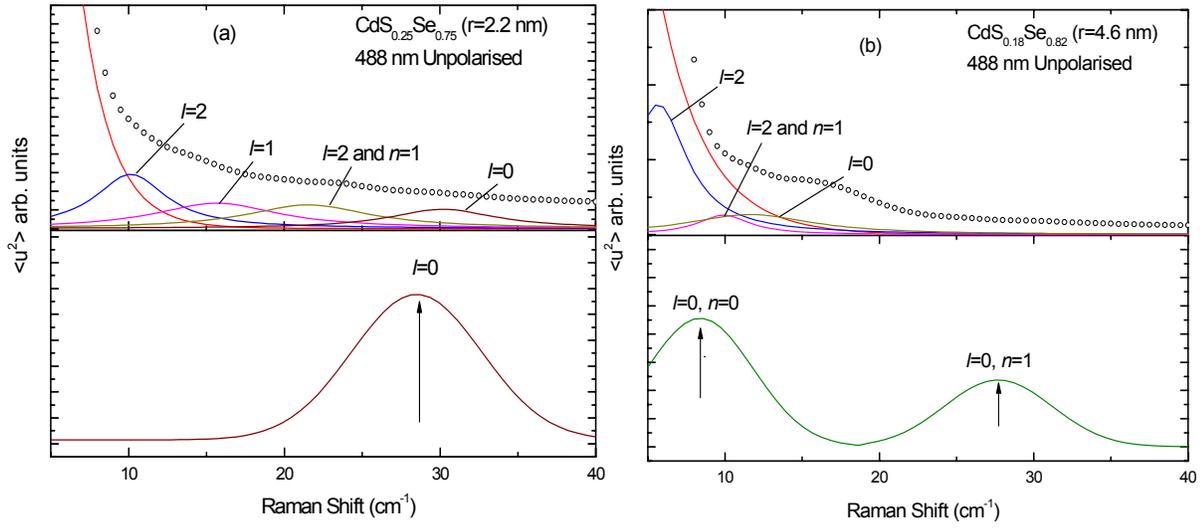

**Figure 1** Gupta *et al.*

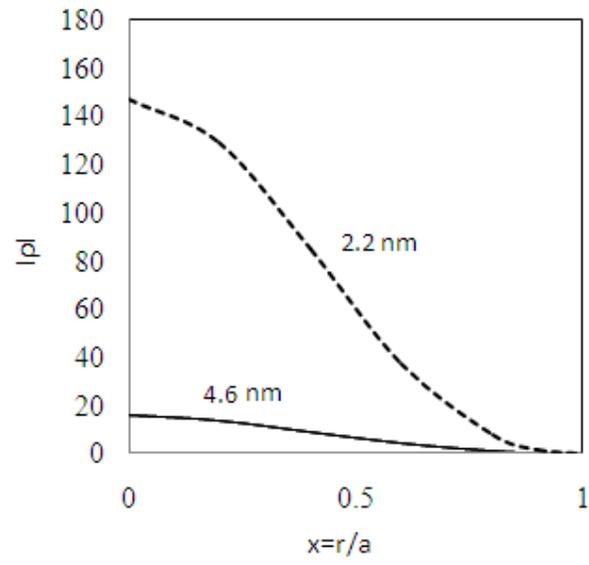

**Figure 2** Gupta *et al.*

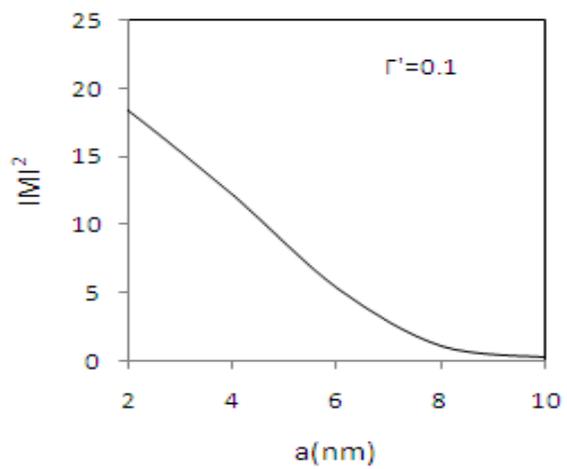

**Figure 3** Gupta *et al.*